\documentclass[aps,prl,
showpacs,superscriptaddress,groupedaddress]{revtex4}  
\usepackage{graphicx}  
\usepackage{dcolumn}   
\usepackage{bm}        
\usepackage{amssymb}   
\usepackage{amsmath}

\begin{document}



\title{Input-Output Formalism for Few-Photon Transport: A Systematic Treatment Beyond Two Photons}
\author{Shanshan Xu}
\affiliation{Department of Physics,
Stanford University, Stanford, California 94305}

\author{Shanhui Fan}
\email{shanhui@stanford.edu} \affiliation{Department of Electrical
Engineering, Ginzton Laboratory, Stanford University, Stanford, California 94305}


\begin{abstract}
We provide a systematic treatment of $N$-photon transport in a waveguide coupled to a local system, using the input-output formalism. The main result of the paper is a general connection between the $N$-photon S matrix and the Green functions of the local system. We also show that the computation can be significantly simplified, by exploiting the connectedness structure of both the S matrix and the Green function, and by computing the Green function using an effective Hamiltonian that involves only the degrees of freedom of the local system. We illustrate our formalism by computing $N$-photon transport through a cavity containing a medium with Kerr nonlinearity, with $N$ up to 3. 
\end{abstract}

\maketitle

\section{\label{sec:level1} I. Introduction}

The capability to create strong photon-photon interaction at a few-photon level in integrated photonic systems is of central importance for quantum information processing. To achieve such a capability, an important approach is to use the so-called waveguide quantum electrodynamics (QED) system, where one confines the photons to a waveguide that is strongly coupled to a local quantum system. Experimentally, the waveguides that have been used for this purpose include optical fibers \cite{adwb}, metallic plasmonic nanowires \cite{amy}, photonic crystal waveguides \cite{lhsj}, and microwave transmission line \cite{wsbf}. The local quantum system typically incorporates a variety of quantum multi-level systems such as actual atoms \cite{adwb}, quantum dots\cite{amy,lhsj}, or microwave qubits \cite{wsbf}, where the strong nonlinearity of these multi-level systems forms the basis for strong photon-photon interactions. These multi-level systems moreover can be embedded in cavity structures to further control their nonlinear properties \cite{bbmbnk,dpa,kp,emf,hbw,lbes}. 

The rapid experimental developments, in turn, have motivated significant theoretical efforts. From a fundamental physics perspective, the photon-photon interaction is characterized by the multi-photon scattering matrix (S matrix). Therefore, a natural objective for theoretical works is to compute such multi-photon S matrix. Moreover, from an engineering perspective, the systems considered here are envisioned as devices that process quantum states. To describe these systems as a device one naturally have to specify its input-output relation. The S matrix, which relates the input and output states, therefore provides a natural basis for device engineering as well.

Motivated by both the physics and engineering considerations as discussed above, a large body of theoretical works have been therefore devoted to the computation the S matrix of various waveguide QED systems \cite{crdl,szgm,gmmtg,sf,sfA,fks,lsb,eks,zb,r,dr,koz,zgb,ll,ll2,sfs,rf,jg,ss,zgb2,rf2,sf2}. These computations, however, are limited in two important aspects:
 
(1)	All of these computations are carried out for a specific local quantum system. In most of these cases, the methods that were used were tailored to the property of the specific system. In the wavefunction approach for S matrix calculation \cite{sf,sfA,lsb,r,dr,koz,ll,ll2,jg,zgb2,szgm}, for example, the ansatz for the wavefunction used is specifically related to the local quantum system. As a result, it has been difficult, from these calculations, to identify the general features of S matrices for waveguide QED systems. 

(2)	With a few exceptions \cite{ss,zgb2}, almost all previous calculations have been carried out for either single or two-photon S matrix. On the other hand, in quantum information processing, there is a strong effort to create and understand highly entangled states with more than two photons \cite{bpd,pdg, zczy}. It is important to understand whether waveguide QED system can be used for such a purpose. Thus, computation of $N$-photon S matrix with $N>2$ is essential.

In this paper, we extend the input-output formalism \cite{fks,eks,rf,rf2,gc} to provide a systematic computation of $N$-photon S matrix for waveguide QED system. The main result of the paper is the relation between the $N$-photon scattering matrix, and the Green function of the local system. We prove this result using only a quantum causality condition, without the need of knowing the specific details of the Hamiltonian of the local system. The main result is therefore generally applicable for a large number of waveguide QED systems with different local quantum systems. We also discuss the general connectedness structure of both the S matrix and the Green function,  which arise from the local nature of the interaction, and show that such connectedness structure can be used to significantly simplify the computation. 

Our work represents a significant step forward in the understanding of waveguide QED system. Our results here highlight some of the universal nature of the properties of these strongly correlated systems that has not been emphasized before. As a computational method, our work here leads to an approach for systematic computation for $N$-photon scattering matrix that is directly applicable to a large number of different systems.

Our work utilizes the input-output formalism developed in standard quantum optics literature. However, our focus here is different. Whereas much of the standard quantum optics literature have focused on computing properties related to an input state that is a coherent state, a thermal state, or a squeezed state, here we focus exclusively on computations for Fock state input.
In general, the transport property of Fock states is qualitatively different from that of the coherent state. As a prominent recent example as developed in the context of Boson sampling problem \cite{bs1,bs2}, it has now been recognized that the $N$-photon Fock state transport in a linear waveguide network is computationally hard \cite{bs3}, even though the transport properties of the same network for coherent states are well known. Similarly, in the system that we are discussing here, while many properties of the system in the presence of a coherent state input can be and have been computed with standard quantum optics tools, much less is known about how to compute transport properties of the same system with a $N$-photon Fock state input.

The paper is organized as follows. 
In Section II we briefly review the input-output formalism and derive the quantum causality condition.
In Section III we prove certain time-ordering relations, which are the key to compute the S matrix of waveguide photons.
In Section IV we prove the connection between  the $N$-photon S matrix and the time-ordered local system's Green function. 
This derivation represents the main result of the paper.  
To further simplify the calculation, we study the connectedness structure of S matrix in Section V.
We also show in Section VI that the system's Green function can be computed with an effective Hamiltonian approach. 
Finally, in Section VII as an example of the application of this formalism, we calculate the exact $N$-photon S matrix up to $N=3$,
when the local system is a cavity containing a medium with Kerr nonlinearity.

\section{II. A brief review of the input-output formalism}
We start with a brief review of the input-output formalism, highlighting only those aspects that will be required for the paper here.
More details can be found in \cite{fks,gc}. Following \cite{fks,gc}, we 
 consider the Hamiltonian of a one-mode waveguide coupled to a local system with finite degrees of freedom ($\hbar=1$):
\begin{equation}\label{H}
H=\int dk\,k\,c_{k}^{\dag}c_k+\xi\int dk\left(c_k^{\dag}a+a^{\dag}c_k\right)+H_{\text{sys}}\,,
\end{equation}
where  $\xi$ is the coupling constant between the waveguide and the system, and is assumed to be frequency independent.  $c_k\, (c_k^{\dag})$ is the annihilation (creation) operator
of the photon state in the waveguide satisfying the
standard commutation relation $[c_k, c_{k'}^{\dag}]=\delta(k-k')$. 
We consider only a narrow range of frequencies, in which the waveguide dispersion relation can be linearized, and the group velocity of the waveguide is taken to be 1.
 $a\,(a^{\dag})$ is one of several possible system operators that are assumed to commute with $c_k,\, c_k^{\dag}$.
In this section we assume $a$ to be arbitrary. In practice $a$ can either be a bosonic operator describing a cavity mode \cite{ll,ll2,sfs,rf,sf2}, or a spin operator for atom-waveguide interaction \cite{sf,fks,eks,r,zgb}.

The aim of the paper is to develop a systematic approach to compute the \textit{N}-photon scattering matrix.   In general, the $N$-photon S matrix is related to the input and output operators by
\cite{fks}  
\begin{equation}\label{NS}
S_{p_1\cdots p_N; k_1 \cdots k_N}=
\left( \prod_{i=1}^{N} \int \frac{d t'_i}{\sqrt{2\pi}}  e^{i p_i t'_i} \prod_{j=1}^{N} \int \frac{d t_j}{\sqrt{2\pi}} e^{-i k_j t_j}\right)
\langle 0| \prod_{i=1}^{N} c_{\text{out}}(t'_i)\prod_{j=1}^{N} c^{\dag}_{\text{in}}(t_j)|0\rangle\,,
\end{equation}
in which we define the input and output operators as:
\begin{eqnarray}\label{dio}
c_{\text{in}}(t)&=&\frac{1}{\sqrt{2\pi}}\int\,dk\, c_k(t_0)\,e^{-ik(t-t_0)}\,,\nonumber\\
c_{\text{out}}(t)&=&\frac{1}{\sqrt{2\pi}}\int\,dk\, c_k(t_1)\,e^{-ik(t-t_1)}\,,
\end{eqnarray}
with $t_0\rightarrow -\infty\,,t_1\rightarrow +\infty$. We note that $c_{\text{in}}(t)$ and $c_{\text{out}}(t)$ consist of Heisenberg operators of waveguide photons at time $-\infty$ and $+\infty$, respectively. They also satisfy the commutation relations 
\begin{equation}\label{cccc}
\left[c_{\text{in}}(t), c_{\text{in}}(t')\right]=\left[c_{\text{out}}(t), c_{\text{out}}(t')\right]=0\,,\,\,\,\,\,
\left[c_{\text{in}}(t), c^{\dag}_{\text{in}}(t')\right]=\left[c_{\text{out}}(t), c^{\dag}_{\text{out}}(t')\right]=\delta(t-t')\,.
\end{equation}

Following the standard procedure \cite{fks,gc}, one can develop the standard input-output formalism that relates $c_{\text{in}}$, $c_{\text{out}}$ and $a$, as:
\begin{equation}\label{ior}
c_{\text{out}}(t)=c_{\text{in}}(t)-i\sqrt{\gamma}\,a(t)\,,
\end{equation}
\begin{equation}\label{ain}
\frac{d a}{dt}=-i\,\sqrt{\gamma}\,\left[a,a^{\dag}\right]\, c_{\text{in}} -i\,\left[a,\,H_{\text{sys}}-i\frac{\gamma}{2}a^{\dag}a\right]\,,
\end{equation}
or
\begin{equation}\label{gain}
\frac{d a}{dt}=-i\,\sqrt{\gamma}\left[a,a^{\dag}\right]\, c_{\text{out}} -i\,\left[a,\,H_{\text{sys}}+i\frac{\gamma}{2}a^{\dag}a\right]\,,
\end{equation}
where $\gamma=2\pi\xi^2$. Integrating (\ref{ain}) and (\ref{gain}) from $t = -\infty$ and $t = \infty$, respectively, result in:
\begin{equation}\label{ie1}
a(t)=a(-\infty) -i\int_{-\infty}^t d\tau\left[a,\,H_{\text{sys}}-i\frac{\gamma}{2}a^{\dag}a\right]-i\,\sqrt{\gamma}\int_{-\infty}^t d\tau \left[a,a^{\dag}\right]\, c_{\text{in}}\,,
\end{equation}
\begin{equation}\label{ie2}
a(t)=a(+\infty) -i\int_{+\infty}^t d\tau\left[a,\,H_{\text{sys}}+i\frac{\gamma}{2}a^{\dag}a\right]-i\,\sqrt{\gamma}\int_{+\infty}^t d\tau \left[a,a^{\dag}\right]\, c_{\text{out}}\,,
\end{equation}
where the integrands in (\ref{ie1}) and (\ref{ie2}) are operators at time $\tau$. 

(\ref{ie1}) and (\ref{ie2}) can be used to prove a quantum causality relation.
When using (\ref{ie1}) to evaluate $a(t)$ or $a^{\dag}(t)$, the integral in (\ref{ie1}) should result in an expression that involves only 
$c_{\text{in}}(\tau)$ and $c^{\dag}_{\text{in}}(\tau)$ with $\tau < t$. Therefore, by the commutation relation (\ref{cccc}) above,  one concludes from (\ref{ie1}) that for $t\leqslant t'$,  
\begin{equation}
\left[a(t),\, I(t')\right]=\left[a(-\infty),\,I(t')\right]\,,\,\,\,\,\,\,\,\left[a^{\dag}(t), \,I(t')\right]=\left[a^{\dag}(-\infty),\,I(t')\right]\,,
\end{equation}
where $I(t')$ is a shorthand notation for the input operators that represent either $c_{\text{in}}(t')$ or $c_{\text{in}}^{\dag}(t')$. On the other hand, the operator $I$ is really a Heisenberg operator at time $-\infty$ as can be seen in (\ref{dio}) above, and hence commute with $a(-\infty)$
and $a^{\dag}(-\infty)$. Therefore, we have 
\begin{equation}\label{cau1}
\left[a(t), \,I(t')\right]  =\left[a^{\dag}(t), I(t')\right] =0\,
\,,\,\,\,\,\,\,\,\text{for}\,\,t\leqslant t'\,,
\end{equation}
Similarly, one can prove
\begin{equation}\label{cau2}
\left[a(t),\,O(t')\right]=\left[a^{\dag}(t),\,O(t')\right]=0\,,
\,\,\,\,\,\,\,\text{for}\,\,t\geqslant t'\,,
\end{equation}
where $O(t')$ is a shorthand notation for the output operators that represent either $c_{\text{out}}(t')$ or $c_{\text{out}}^{\dag}(t')$,
by utilizing (\ref{ie2}) and the fact that the output operators are really Heisenberg operators at time $+\infty$. 
Following \cite{gc}, we refer to (\ref{cau1}) and (\ref{cau2}) as the \textit{quantum causality condition}. The operator $a(t)$, which characterizes the physical field in the local system, depends only on the input field $c_{\text{in}}(\tau)$ with $\tau<t$, and generate only output field $c_{\text{out}}(\tau)$ with $\tau>t$. 

\section{III Relation involved time-ordered product}
Having reviewed some of the basic aspects of the input-output formalism, we now proceed to compute the \textit{N}-photon S matrix as defined in (\ref{NS}). 
For this purpose, we first consider some of the properties of a time-ordered product involving 
$a$ and the input or output operators. We note that:
\begin{eqnarray}\label{tor}
{\cal{T}} a(t)I(t')&=&a(t)I(t')\,,\\
{\cal{T}} a(t)O(t')&=&O(t')a(t)\label{tor2}\,.
\end{eqnarray}
Take (\ref{tor}) as an example, by definition, ${\cal{T}} a(t)I(t')=a(t)I(t')$ for $t\geqslant t'$; When $t<t'$, by  the quantum causality condition of (\ref{cau1}),
we have ${\cal{T}} a(t)I(t')=I(t')a(t)=a(t)I(t')$, completing the proof. (\ref{tor2}) can be proved similarly. 

More generally, we have the following relation regarding the time-ordered product:
\begin{eqnarray}\label{timeorder}
{\cal{T}}\prod_{i,j} a(t_i)I(t'_j)&=&\left[{\cal{T}}\prod_{i} a(t_i)\right]\cdot\left[{\cal{T}}\prod_{j}I(t'_j)\right]\,,\\
{\cal{T}}\prod_{i,j} a(t_i)O(t'_j)&=&\left[{\cal{T}}\prod_{j}O(t'_j)\right]\cdot\left[{\cal{T}}\prod_{i} a(t_i)\right]\label{timeorder2}\,,
\end{eqnarray}
where bracket is used to indicate the range over which the time-ordering is being applied. 

(\ref{timeorder}) and (\ref{timeorder2}) can be proved in a similar way. Here we show only the proof of (\ref{timeorder}). 
The proof of (\ref{timeorder}) can be constructed from induction with respect to the number of operators.
The base case is already proved in (\ref{tor}). Now suppose (\ref{timeorder}) holds for all cases involving a total number of $N$ operators of $a$ and $I$.
Consider a time-ordered product involving $N+1$ operators, 
if the operator with the largest time label is $a(t_{\text{max}})$,
\begin{eqnarray}
{\cal{T}}\prod_{I,j} a(t_I)I(t'_j)=a(t_{\text{max}})\left[{\cal{T}}\prod_{i,j} a(t_i)I(t'_j)\right]
=a(t_{\text{max}})\left[{\cal{T}}\prod_{i} a(t_i)\right]\cdot\left[{\cal{T}}\prod_{j}I(t'_j)\right]
=\left[{\cal{T}}\prod_{I} a(t_I)\right]\cdot\left[{\cal{T}}\prod_{j}I(t'_j)\right]\,,\nonumber\\
\end{eqnarray}
where the definition of time-ordered product is used in the first and last steps and the induction hypothesis is used in the second step.
On the other hand, if the operator with the largest time label is $I(t_{\text{max}})$, we have
\begin{eqnarray}
{\cal{T}}\prod_{i,J} a(t_i)I(t'_J)&=&I(t_{\text{max}})\left[{\cal{T}}\prod_{i,j} a(t_i)I(t'_j)\right]
= I(t_{\text{max}})\left[{\cal{T}}\prod_{i} a(t_i)\right]\cdot\left[{\cal{T}}\prod_{j}I(t'_j)\right]\nonumber\\
&=& \left[{\cal{T}}\prod_{i} a(t_i)\right]\cdot I(t_{\text{max}})\left[{\cal{T}}\prod_{j}I(t'_j)\right]
=\left[{\cal{T}}\prod_{i} a(t_i)\right]\cdot\left[{\cal{T}}\prod_{J}I(t'_J)\right]\,,\nonumber\\
\end{eqnarray}
where we use the induction hypothesis in the second step and the commutation relation (\ref{cau1}) in the third step. Therefore, (\ref{timeorder}) holds for $N+1$ operators, completing the proof.

\section{IV \textit{N}-photon S matrix}
Using  (\ref{timeorder}) and (\ref{timeorder2}), we now compute the $N$-photon S matrix defined in (2). We first evaluate its Fourier transformation in the time domain
\begin{equation}\label{NSt}
S_{t'_1\cdots t'_N; t_1\cdots t_N}\equiv\langle 0| \prod_{i=1}^{N} c_{\text{out}}(t'_i)\prod_{j=1}^{N} c^{\dag}_{\text{in}}(t_j)|0\rangle\,.
\end{equation}
From now on, we refer both (\ref{NS}) and (\ref{NSt}) as S matrix. The computation is as follows:
\begin{eqnarray}
&&S_{t'_1\cdots t'_N; t_1\cdots t_N} \nonumber\\
&=&  \langle 0|\left[ {\cal{T}} \prod_{i=1}^N c_{\text{out}}(t_i')\right] \cdot \prod_{j=1}^N c_\text{{in}}^{\dag}(t_j) |0\rangle \nonumber\\
&=& \langle 0|  \left[{\cal{T}}  \prod_{i=1}^N \left(c_\text{{in}}(t_i')-i\sqrt{\gamma}\,a(t_i') \right)\right] \cdot \prod_{j=1}^N c_\text{{in}}^{\dag}(t_j) |0\rangle\nonumber\\
&=&  \sum_{M=0}^N \sum_{B_M} (-i\sqrt{\gamma})^M   \langle 0|  \left[{\cal{T}} \prod_{s=1}^{M} a(t_{B_M(s)}')\right] \cdot \left[{\cal{T}} \prod_{m=1}^{N-M} c_\text{{in}}(t_{B_M^c(m)}')\right]  \cdot   \prod_{j=1}^N c_\text{{in}}^{\dag}(t_j)  |0\rangle\nonumber\\
&=& \sum_{M=0}^N  (-i\sqrt{\gamma})^M   \sum_{B_M,D_M} \langle 0|  {\cal{T}}\prod_{s=1}^{M} a(t_{B_M(s)}')  \prod_{r=1}^{M} c_\text{{in}}^{\dag}(t_{D_M(r)}) |0\rangle  \sum_P \prod_{m=1}^{N-M} \delta\left(t_{B_M^c(m)}'-t_{PD_M^c(m)}\right) \nonumber\\
&=& \sum_{M=0}^N  (-i\sqrt{\gamma})^M   \sum_{B_M,D_M} \langle 0|  {\cal{T}} \prod_{s=1}^{M} a(t_{B_M(s)}')\prod_{r=1}^{M} \left(c_{\text{out}}^{\dag}(t_{D_M(r)}) - i\sqrt{\gamma}\, a^{\dag}(t_{D_M(r)}) \right) |0\rangle\sum_P \prod_{m=1}^{N-M} \delta\left(t_{B_M^c(m)}'-t_{PD_M^c(m)}\right) \nonumber\\
&=& \sum_{M=0}^N (-\gamma)^{M}  \sum_{B_M,D_M} \langle 0|  {\cal{T}} \prod_{s=1}^{M} a(t_{B_M(s)}')  \prod_{r=1}^{M}  a^{\dag}(t_{D_M(r)})|0\rangle \sum_P \prod_{m=1}^{N-M} \delta\left(t_{B_M^c(m)}'-t_{PD_M^c(m)}\right).
\end{eqnarray}

In the first step above, we introduce the time-ordering operation for the product of $c_\text{{out}}$ operators since these operators commute with one another. 

In the second step, we use the input-output relation (\ref{ior}). 

In the third step, we expand the product in the bracket.  In this derivation, for a given subset $B$ of $\{1...N\}$, we use $B^c$ to represent its corresponding complementary subset, and $B(s)$ to represent its $s$-th element. The summation is over all subsets of  $\{1...N\}$. $B_M$ is a subset with $M$ elements. We use the time-ordering relation (\ref{timeorder}) to place all $c_\text{{in}}$ operators to the right of the $a$ operators.  

In the fourth step, we first remove the time-ordering operation for the product of $c_\text{{in}}$ and then contract the $c_\text{{in}}$ operators with the $c_{\text{in}}^{\dag}$ operators. Since there are $N-M$ $c_\text{{in}}$  and $N$ $c_\text{{in}}^{\dag}$ operators, in each term in the resulting summation we select $M$ $c_\text{{in}}^{\dag}$ operators as indexed by a $M$-element subset $D_M$, and perform a full contraction using the remaining $N-M$ $c_\text{{in}}^{\dag}$ operators.  Such contraction results in a summation over all possible permutations $P$ of $D_M^c$. Finally, we restore the time-ordering operation on the product of $M$ $c_\text{{in}}^{\dag}$ operators. The use of (\ref{timeorder}) then results in a time-ordered product of all the $a$ and $c_\text{{in}}^{\dag}$ operators. 

In the fifth step, we again use the input-output relation (\ref{ior}).

In the last step, we expand the product involving $a^{\dag}$, and then apply (\ref{timeorder2}) to each term in the product expansion. For every term that contains at least one $c_{\text{out}}^{\dag}$ operators, the use of (\ref{timeorder2}) resulting in such output operators being placed on the left-most positions of the operator product, and hence such a term vanishes. 

Therefore, we obtain the first main result of the paper:
\begin{equation}\label{central}
S_{t'_1\cdots t'_N; t_1\cdots t_N} =\sum_{M=0}^N (-\gamma)^{M}  \sum_{B_M,D_M} \langle 0|  {\cal{T}} \prod_{s=1}^{M} a(t_{B_M(s)}')  \prod_{r=1}^{M}  a^{\dag}(t_{D_M(r)})|0\rangle \sum_P \prod_{m=1}^{N-M} \delta\left(t_{B_M^c(m)}'-t_{PD_M^c(m)}\right).
\end{equation}

We define the time-ordered 2M-point Green function 
\begin{equation}\label{Gs}
G(t'_1\cdots t'_M; t_1\cdots t_M)\equiv (-\gamma)^{M}  \langle 0|  {\cal{T}}a(t'_1)\cdots a(t'_M)a^{\dag}(t_1)\cdots a^{\dag}(t_M)|0\rangle\,,
\end{equation}
and its Fourier transformation
\begin{equation}\label{NGP}
G(p_1\cdots p_M; k_1\cdots k_M)\equiv
\left( \prod_{i=1}^M \int \frac{d \,t'_i}{\sqrt{2\pi}} e^{i p_i t'_i} \prod_{j=1}^M \int \frac{d\, t_j}{\sqrt{2\pi}} e^{-i k_j t_j}\right)
G(t'_1\cdots t'_M; t_1\cdots t_M)\,.
\end{equation} 
From (\ref{central}), the \textit{N}-photon S matrix (\ref{NS}) in the frequency domain is  
\begin{equation}\label{cenm}
S_{p_1\cdots p_N; k_1\cdots k_N}=\sum_{M=0}^N \sum_{B_M,D_M} G\left(p_{B_M}; k_{D_M}\right) \sum_P \prod_{m=1}^{N-M} \delta\left(p_{B_M^c(m)}-k_{PD_M^c(m)}\right)\,,
\end{equation}
where we use the shorthand notation $p_{B_M}\equiv \{p_i|i\in {B_M}\}$ and $k_{D_M}\equiv \{k_i|i\in {D_M}\}$. 
In (\ref{cenm}), $ k_{PD_M^c(m)}$ represents the frequencies of the incoming photons that bypass the local system. These photons do not change their frequencies, as signified by the   $\delta$ functions in (\ref{cenm}). Whereas $k_{D_M}$ represents the frequencies of the incoming photons that enter and are scattered by the local system.  

\begin{figure}
\includegraphics
[width=0.8\textwidth] {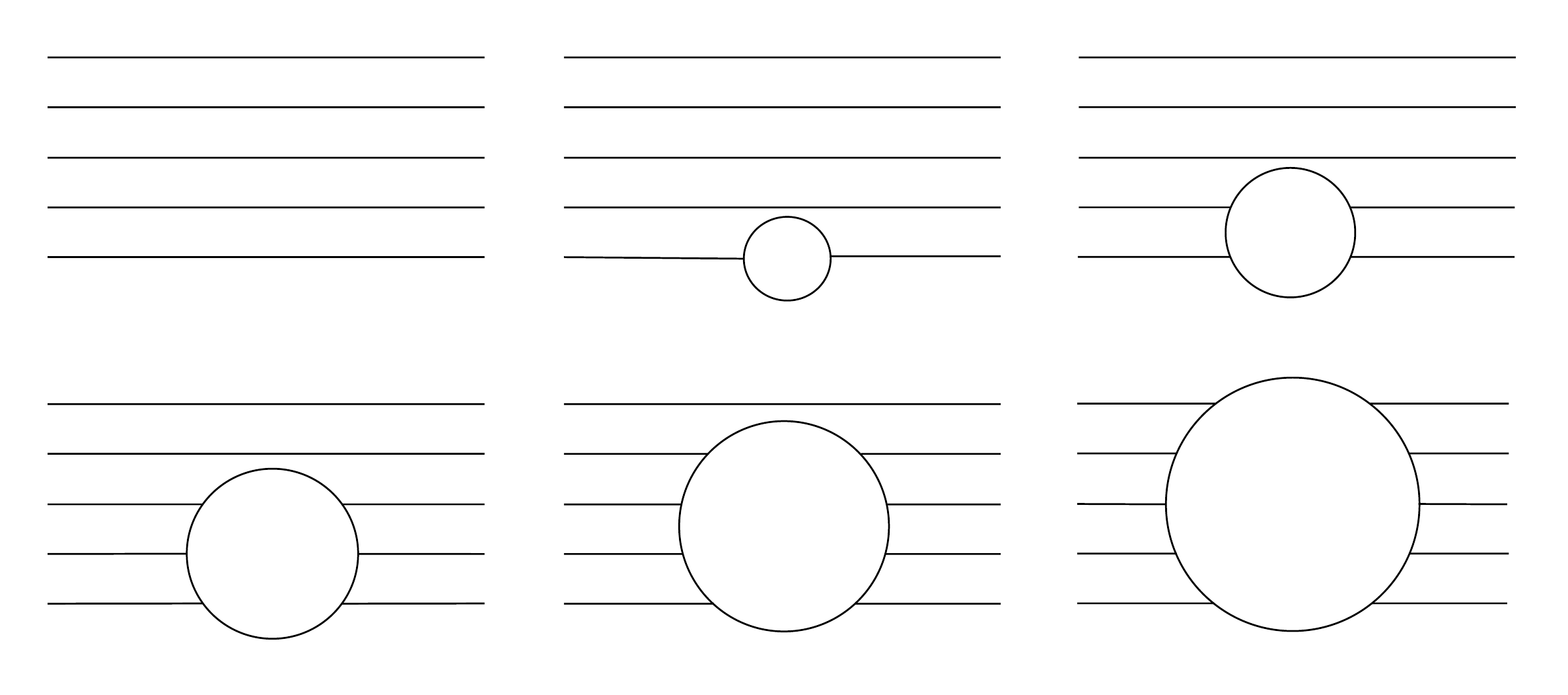} \caption{
The diagrammatic representation of all six classes of terms that arise in the computation of S matrix for $N=5$ photons.
} \label{fig1}
\end{figure}

\begin{figure}
\includegraphics
[width=0.8\textwidth] {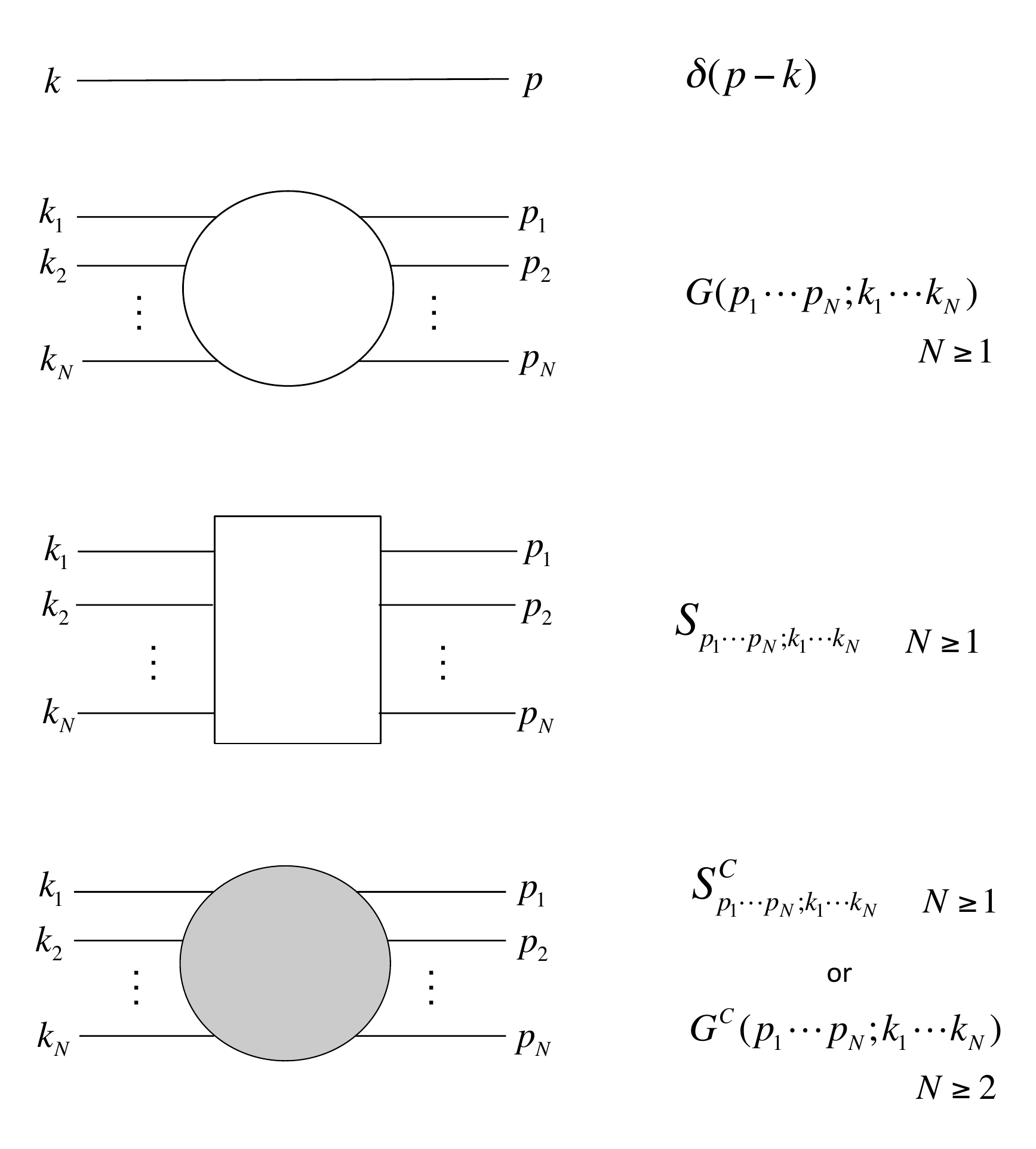} \caption{
Various diagrams used in the paper. The corresponding mathematical definition is shown on the right. The legs to the left (right) represent input (output) momenta. There is no distinction in the ordering among the input or output momenta. 
} \label{fig2}
\end{figure}

Each term in (\ref{cenm}) can be represented diagrammatically. For the case with $N=5$, for example, we plot all different classes of diagrams in Fig.\ref{fig1}. (We summarize the definitions of all diagrams used in the paper in Fig.\ref{fig2}). The summation in (\ref{cenm}) then represents the summation of all such diagrams as shown in Fig.\ref{fig1}, each diagram containing a single Green function part, and the rest are $\delta$ functions.  

Our main result (\ref{cenm}) directly reduces the computation of the $N$-photon S matrix to the calculation of the Green function of the \textit{local system}. The result here is related to, but not identical to, the LSZ reduction approach as discussed in \cite{sfs}. In \cite{sfs}, the $N$-photon S matrix is first reduced with the LSZ reduction to the calculation of the Green function of the \textit{waveguide} photons, which is then related to the Green function of the local system after integrating out of the waveguide photon fields. In contrast, in our derivation the reduction to the Green function of the local system is obtained directly with input-output formalism, which represents a more direct approach. Also, the proof here is very general. It uses only the quantum causality relation without any need for knowing the details of the Hamiltonian of the local system, which again points to the power of the input-output formalism.

\section{V The connectedness structure of \textit{N}-photon S matrix}
Our main result (\ref{central}) and (\ref{cenm}) reduce the problem of computing $N$-photon S matrix, to the calculation of Green functions of the local system. Such a calculation, moreover, can be significantly simplified exploiting the connectedness structure of the S matrix and the Green function. 

Since the system is time-translation invariant, $G(p_1,\cdots,p_N;k_1,\cdots,k_N)$ must be proportional to $\delta\left(\sum_{i=1}^N p_i - \sum_{i=1}^N k_i\right)$.
In general, $G(p_1,\cdots,p_N;k_1,\cdots,k_N)$ can be expressed as a sum of various terms containing products of several $\delta$ functions \cite{vpn}. Among all these terms, we define the term that contains only $\delta\left(\sum_{i=1}^N p_i - \sum_{i=1}^N k_i\right) $ and no other $\delta$ functions as the \textit{connected}  Green function, $G^C(p_1,\cdots,p_N;k_1,\cdots,k_N)$.  Similarly, as can be seen in (\ref{cenm}), the S matrix can also be organized as summing over various terms containing products of several $\delta$ functions. Among all these terms, we can again define the term that contains only $\delta \left(\sum_{i=1}^N p_i - \sum_{i=1}^N k_i\right)$ and no other $\delta$ function as the 
\textit{connected} part of the S matrix, $S^C_{p_1\cdots p_N;k_1\cdots k_N}$. 

\begin{figure}
\includegraphics
[width=0.8\textwidth] {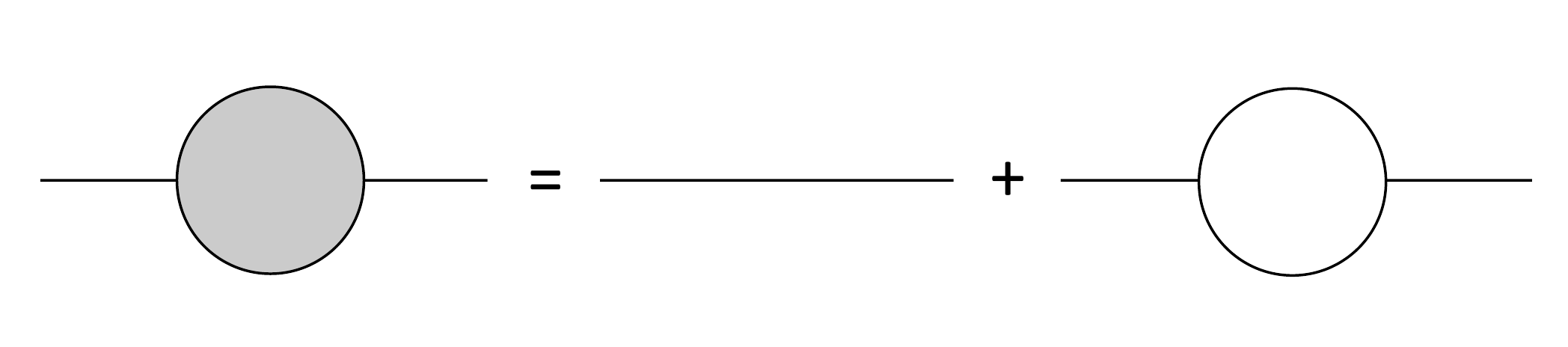} \caption{
The single-photon S matrix is the sum of a $\delta$ function and a two-point Green function. 
} \label{fig3}
\end{figure}

Our main result (\ref{cenm}) then immediately implies that for $N>1$
\begin{equation}\label{SCGC}
S^C_{p_1\cdots p_N;k_1\cdots k_N} = G^C(p_1,\cdots,p_N;k_1,\cdots,k_N)\,,
\end{equation}
and for $N=1$ 
\begin{equation}\label{S1G1}
S_{p_1;k_1} = \delta(p_1-k_1)+G^C(p_1;k_1)\,.
\end{equation}
Therefore, for $N>1$, we use the same diagrammatic representation for $S^C$ and $G^C$ (Fig.\ref{fig2}). For $N=1$, the diagrammatic representation of $S^C$ is included in Fig.\ref{fig2} and the result of (\ref{S1G1}) can then be represented in Fig.\ref{fig3}.

Moreover, it is known that a Green function in general can be cluster-decomposed as \cite{vpn, QFT2}:
\begin{equation}\label{dGC}
G(p_1,\cdots, p_N;k_1,\cdots, k_N) = \sum_{\cal{B}} \sum_P \prod_{i=1}^{M_{\cal{B}}} G^C\left(p_{{\cal{B}}_i};k_{{\cal{B}}P_i}\right)
\end{equation}
where $\cal{B}$ is a partition of an ordered list $\{1,2,\cdots, N\}$ into a collection of subsets. (For example, for a list $\{1,\cdots, 5\}$, a partition results in $\{1\}, \{2,4\}, \{3,5\}$.) ${\cal{B}}P$  is the same partition as $\cal{B}$ but acts on the permutated ordered list 
$\{P(1), P(2), \cdots, P(N)\}$. (For example, the same $\cal{B}$ in the example above, acting on the permutation $\{5,3,4,2,1\}$ would result in $\{5\}, \{32\}, \{41\}$.) 
$M_{\cal{B}}$ is the number of subsets and ${\cal{B}}_i$, ${\cal{B}}P_i$ are the respective $i$-th subsets. We also use the shorthand notations $p_{{\cal{B}}_i}\equiv\{p_j| j\in {\cal{B}}_i\}$ and $k_{{\cal{B}}P_i}\equiv \{k_j|j \in {{\cal{B}}P_i}\}$. For a given partition, we only sum over all the distinct permutations under the exchange symmetries in each sublist. 
Graphically, (\ref{dGC}) states that the Green function can be decomposed into a summation of all possible products of the connected Green functions. 

\begin{figure}
\includegraphics
[width=1.0\textwidth] {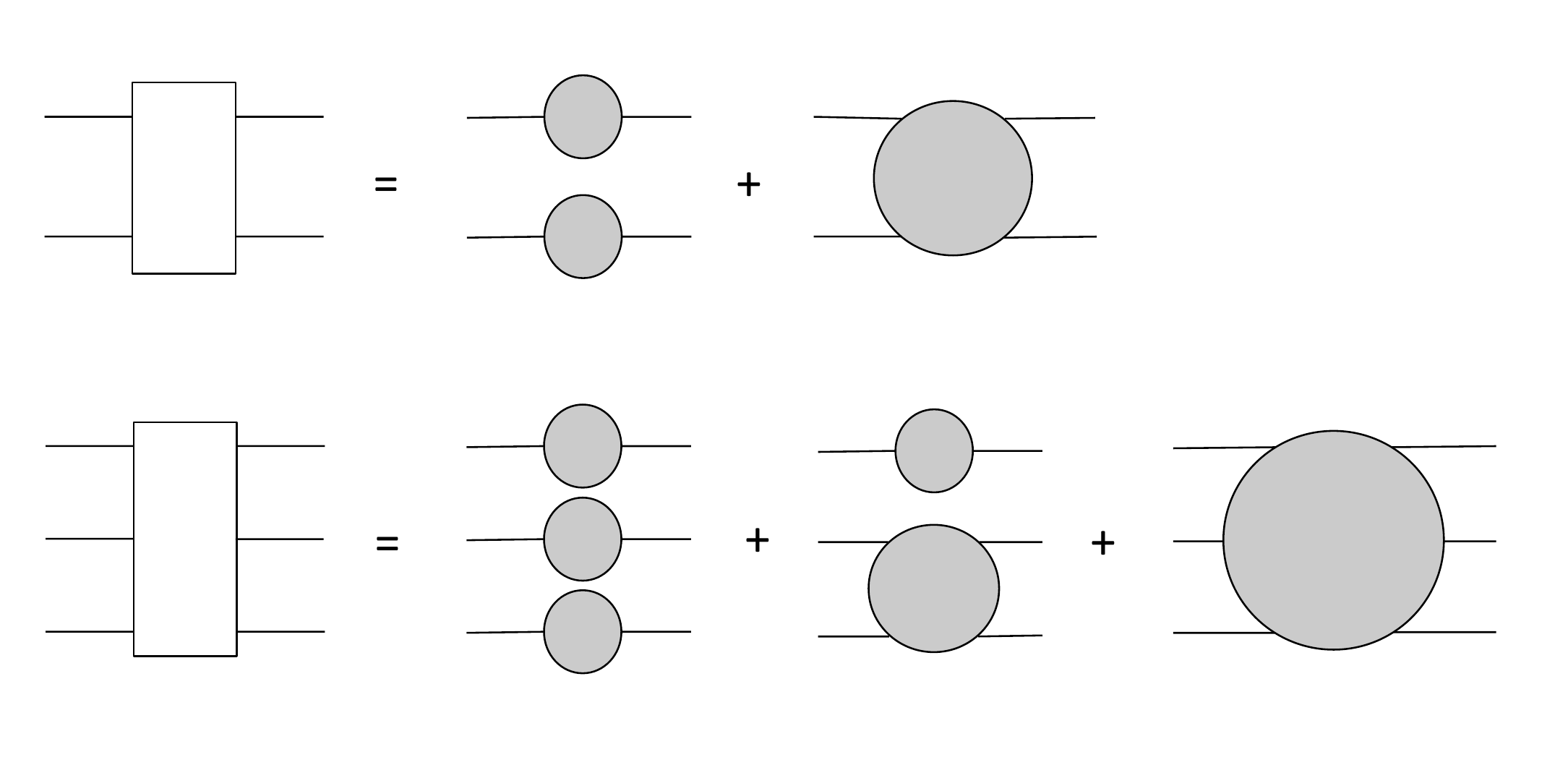} \caption{
The cluster decomposition of two- and three-photon S matrices.
} \label{fig4}
\end{figure}

By (\ref{SCGC}), (\ref{S1G1}) and (\ref{dGC}), we prove in the appendix that our main result (\ref{cenm}) implies similar cluster-decomposition properties for the S matrix of the system \cite{wc,QFT,QFT2}: 
\begin{equation}\label{dSC}
S_{p_1\cdots p_N;k_1\cdots k_N} = \sum_{\cal{B}} \sum_P \prod_{i=1}^{M_{\cal{B}}} S^C_{p_{{\cal{B}}_i};k_{{\cal{B}}P_i}}\,.
\end{equation}
Below, we give three simplest examples of (\ref{dSC}) when $N=1,2,3$, respectively: 
\begin{equation}\label{1t}
S_{p_1; k_1}=\delta(p_1-k_1)+G(p_1; k_1)=S^C_{p_1;k_1}\,.
\end{equation}

\begin{equation}\label{2tC}
S_{p_1p_2; k_1k_2}=S_{p_1; k_1}S_{p_2; k_2}+S_{p_1; k_2}S_{p_2; k_1}+S^C(p_1,p_2; k_1,k_2)\,.
\end{equation}
\begin{eqnarray}\label{3tC}
S_{p_1p_2p_3; k_1k_2k_3}&=&\sum_{P}S_{p_1;k_{P_1}}S_{p_2;k_{P_2}}S_{p_3;k_{P_3}}\nonumber\\
&&+\sum_{P_{\text{even}}}\left[S^C_{p_1, p_2;k_{P_1},k_{P_2}}S_{p_3;k_{P_3}}+S^C_{p_2, p_3;k_{P_2},k_{P_3}}S_{p_1;k_{P_1}}+S^C_{p_1, p_3;k_{P_1},k_{P_3}}S_{p_2;k_{P_2}}\right]\nonumber\\
&&+S^C(p_1,p_2,p_3; k_1,k_2,k_3)\,.
\end{eqnarray}
(\ref{2tC}) and (\ref{3tC}) are represented diagrammatically in Fig.\ref{fig4}. Using such diagrammatic representation, $N$ photon S matrix can be straightforwardly decomposed. The results here thus reduce the computation of the $N$-photon S matrix, to the evaluation of the connected $2N$-point Green function of the local system.  

\section{VI Compute system's Green function with the effective hamiltonian for the system}
In this section, we will prove that the Green function of (\ref{Gs}) can be computed using the effective Hamiltonian of the local system. The main result in this section is
\begin{equation} \label{id}
G({t'_1\cdots t'_N; t_1\cdots t_N})=\widetilde{G}({t'_1\cdots t'_N; t_1\cdots t_N})\,,
\end{equation}
where
\begin{equation}\label{aGs}
\widetilde{G}({t'_1\cdots t'_N; t_1\cdots t_N})\equiv  (-\gamma)^N \langle 0| {\cal{T}}\,\widetilde{a}(t'_1) \cdots \widetilde{a}(t'_N)\widetilde{a^{\dag}}(t_1)\cdots \widetilde{a^{\dag}}(t_N) |0\rangle\,,
\end{equation}
with operators 
\begin{equation}\label{effa}  
\widetilde{a}(t)=e^{i H_{\text{eff}} t}\, a\, e^{-i H_{\text{eff}} t}\,,\,\,\,\,\,\,\,\,\widetilde{a^{\dag}}(t)=e^{i H_{\text{eff}} t} \,a^{\dag}\, e^{-i H_{\text{eff}} t}\,,
\end{equation}
where
\begin{equation}\label{eff}
H_{\text{eff}}\equiv H_{\text{sys}}-i\frac{\gamma}{2}a^{\dag}a\,
\end{equation}
is the effective Hamiltonian for the system. With the identity (\ref{id}), the computation of the Green function is simplified since no operators of waveguide photons are involved.  
We only need to solve a system which has a finite, and typically small, number of degrees of freedom. 

(\ref{id}) can be proved in the path integral formulation. The proof is summarized as:
\begin{eqnarray}\label{GpG}\label{gpi}
G({t'_1\cdots t'_N; t_1\cdots t_N})&=&(-\gamma)^N\frac{\int {\cal{D}}\left[c_k,c^{*}_k, a,a^*\right]a(t'_1) \cdots a(t'_N)a^{*}(t_1)\cdots a^{*}(t_N) e^{i\int dt L}}{\int {\cal{D}}\left[c_k,c^{*}_k,a, a^* \right] e^{i\int dt L}}\nonumber\\
&=&(-\gamma)^N\frac{\int {\cal{D}}\left[a,a^*\right]a(t'_1) \cdots a(t'_N)a^{*}(t_1)\cdots a^{*}(t_N) e^{i\int dt L_{\text{eff}}}}{\int {\cal{D}}\left[a, a^* \right] e^{i\int dt L_{\text{eff}}}}\nonumber\\
&=&\widetilde{G}({t'_1\cdots t'_N; t_1\cdots t_N})\,.
\end{eqnarray}

The first step in (\ref{gpi}) follows \cite{ps} to express Green function (\ref{Gs}) by path integral. Here for simplicity, we assume the local system is characterized only by $a$ and $a^*$. $L$ is 
the Lagrangian associated with the full Hamiltonian (\ref{H}):
\begin{equation} 
L=\int dk \,c_k^{*}(i\partial_t-k)c_k-\xi\int dk\left(c_k^{*}a+a^{*}c_k\right)+L_{\text{sys}}\,,
\end{equation}
in which $L_{\text{sys}}$ is the local system's Lagrangian obtained by Legendre transformation on the system's Hamiltonian $H_{\text{sys}}$.

The second step in (\ref{gpi}) is the key of the proof. As was done in the standard approach involving generating functional,  we introduce the propagator of free waveguide photon 
\begin{equation}
G^{(0)}_k(t-t')\equiv\int \frac{d \omega}{2\pi}\,e^{-i\omega
(t-t')}\frac{i}{\omega-k+i 0^+}
\end{equation}  
and integrate out the waveguide degrees of freedom:
\begin{eqnarray}
\int {\cal{D}}\left[c_k, c_k^* \right] e^{i\int dt L}&=& e^{i\int dt L_{\text{sys}}}\int {\cal{D}}\left[c_k, c_k^* \right] e^{i\int dt \int dk \left[c_k^{*}(t)(i\partial_t-k)c_k(t)-\xi c_k^{*}(t)a(t)-\xi a^{*}(t)c_k(t)\right]}\nonumber\\
&=&e^{i\int dt L_{\text{sys}}}\int {\cal{D}}\left[c_k, c_k^* \right] e^{-\xi^2\int dt \int dt' a^*(t)W(t-t')a(t')}\,,
\end{eqnarray} 
where 
\begin{eqnarray}
W(t-t')&=&\int dk\, G^{(0)}_k(t-t')=i\int \frac{d \omega}{2\pi}\,e^{-i\omega
(t-t')}\int dk\left[\frac{{\cal{P}}}{\omega-k}-i\pi\delta(\omega-k)\right]\nonumber\\
&=&\pi \int \frac{d \omega}{2\pi}\,e^{-i\omega
(t-t')}=\pi\delta(t-t')\,.
\end{eqnarray}
As a result, we obtain the effective Lagrangian
\begin{equation} \label{Leff}
L_{\text{eff}}= L_{\text{sys}}+i\pi\xi^2 a^{*}a=L_{\text{sys}}+i\frac{\gamma}{2} a^{*}a\,.
\end{equation}
Here, the imaginary part of the effective Lagrangian arises since the waveguide degrees of freedom that we are integrating out forms a continuum. 

In the last step of (\ref{gpi}),
the path integral to the system's degrees of freedom with the effective Lagrangian (\ref{Leff})  corresponds exactly to the  alternative Green function (\ref{aGs}) in the Heisenberg picture \cite{ps}. The effective Hamiltonian (\ref{eff}) is obtained from the effective Lagrangian (\ref{Leff}) by Legendre transformation. 

Since in practice, all evaluations of the Green function will be carried out using the effective Hamiltonian (\ref{eff}), in the following we will no longer make the distinction between the Green function of system defined in the full coupled Hamiltonian versus the effective Hamiltonian, i.e. we will no longer make distinction between the left and right hand side of (\ref{id}). 
In what follows, the time evolution of all system operators is considered as that in (\ref{effa}).

(\ref{id}) allows us to compute the full Green function of the local system. To compute the S matrix, we only need the connected part of the Green function. In practice, the connected part of the Green function may actually be obtained in a simpler fashion without the need to evaluate and perform a cluster decomposition of the full Green function. This will be illustrated in the example below.

\section{VII Example: three-photon S matrix with Kerr nonlinear cavity}
As an example of the application of the formalism developed in this paper, we compute the S matrix of three-photon transport in a
single-mode waveguide side-coupled to a ring resonator incorporating Kerr nonlinear media. The full Hamiltonian has the same form as (\ref{H}) with the specific form of $H_{\text{sys}}$:
\begin{equation}
H_{\text{sys}}=\omega_c \,a^{\dag}a+\frac{\chi}{2}\,a^{\dag}a^{\dag}aa\,
\end{equation} 
where $a$ is the annihilation operator of cavity photon satisfying the standard commutator relation $\left[a, a^{\dag}\right]=1$. Let $\alpha\equiv \omega_c-i\frac{\gamma}{2}$,
the effective Hamiltonian (\ref{eff}) in this case is 
\begin{equation}
H_{\text{eff}}=\alpha \,a^{\dag}a+\frac{\chi}{2}\,a^{\dag}a^{\dag}aa\,,
\end{equation}
and can be diagonalized as 
\begin{equation}\label{eigen}
H_{\text{eff}}|n\rangle=\left[\alpha n+\frac{\chi}{2}n(n-1)\right]|n\rangle\,.
\end{equation}

Use the formalism as described above, we now compute the S matrix for up to three photons. From (\ref{1t})-(\ref{3tC}) we only need the connected S matrix for up to three photons, which is the focus of the calculation here. 
We consider single photon S matrix first.  Using (\ref{id}), the two-point Green function is computed as 
\begin{eqnarray}\label{G1}
G(t'; t)=-\gamma\sum_{n}\langle 0|{a}(t')|n\rangle\langle n|{a^{\dag}}(t)|0\rangle\theta(t'-t)
=-\gamma\,e^{-i \alpha(t'-t)}\theta(t'-t)\,,
\end{eqnarray}
where a completeness set of states of the system is inserted. To obtain the final result, we note that only the single photon state contributes to the summation. 
The single photon S matrix is then:
\begin{equation}\label{sS}
S_{p;k}=\frac{k-\omega_c-i\gamma/2}{k-\omega_c+i\gamma/2}\delta(p-k)=\left[1+s_k\right]\delta(p-k)\,,
\end{equation}
where, for later convenience, we defined 
\begin{equation}
s_k\equiv -\gamma\frac{i}{k-\alpha}\,.
\end{equation} 

To compute the connected two-photon S matrix, we first compute the four-point Green function $G_2(t'_1,t'_2;t_1,t_2)$.
Depending on the values of the four time labels, the time ordering operation would give rise non-zero terms that can 
be classified into two types: $\langle {a}{a^{\dag}}{a}{a^{\dag}}\rangle$ and $\langle {a}{a}{a^{\dag}}{a^{\dag}}\rangle$, i.e.
\begin{equation}
G(t'_1,t'_2; t_1,t_2)\equiv \sum_{j=1}^2 G^{(j)}(t'_1,t'_2; t_1,t_2)\,,
\end{equation}
with 
\begin{eqnarray}
G^{(1)}(t'_1,t'_2; t_1,t_2)&=&(-\gamma)^2\sum_{P,Q}\langle 0| {a}(t'_{Q_1}){a^{\dag}}(t_{P_1}){a}(t'_{Q_2}){a^{\dag}}(t_{P_2})|0\rangle\theta(t'_{Q_1}-t_{P_1})\theta(t_{P_1}-t'_{Q_2})\theta(t'_{Q_2}-t_{P_2})\,,\label{cor21}\\
G^{(2)}(t'_1,t'_2; t_1,t_2)&=&(-\gamma)^2\sum_{P,Q}\langle 0| {a}(t'_{Q_1}){a}(t'_{Q_2}){a^{\dag}}(t_{P_1}){a^{\dag}}(t_{P_2})|0\rangle\theta(t'_{Q_1}-t'_{Q_2})\theta(t'_{Q_2}-t_{P_1})\theta(t_{P_1}-t_{P_2})\,,\label{cor22}
\end{eqnarray}
where both $P$ and $Q$ are permutations over indices $\{1,2\}$.
We calculate each term by inserting the complete sets of eigenstates, which results in:
\begin{eqnarray}
\langle 0| {a}(t'_1){a^{\dag}}(t_1){a}(t'_2){a^{\dag}}(t_2)|0\rangle
&=&\sum_{m, n,l}\langle 0| {a}(t'_1)|m\rangle\langle m|{a^{\dag}}(t_1)|n\rangle
\langle n | {a}(t'_2)| l\rangle\langle l |{a^{\dag}}(t_2)|0\rangle
\nonumber\\
&=&\langle 0| {a}(t'_1)|1\rangle\langle 1|{a^{\dag}}(t_1)|0\rangle
\langle 0 | {a}(t'_2)| 1\rangle\langle 1 |{a^{\dag}}(t_2)|0\rangle
=e^{-i\alpha (t'_1-t_1)}e^{-i\alpha(t'_2-t_2)}\,,
\end{eqnarray}
and
\begin{eqnarray}
\langle 0| {a}(t'_1){a}(t'_2){a^{\dag}}(t_1){a^{\dag}}(t_2)|0\rangle
&=&\langle 0| {a}(t'_1)|1\rangle\langle 1|{a}(t'_2)|2\rangle\langle 2 |{a^{\dag}}(t_1)| 1\rangle\langle 1 |{a^{\dag}}(t_2)|0\rangle
\nonumber\\
&=&2e^{-i\alpha(t'_1-t_2)}e^{-i(\alpha+\chi)(t'_2-t_1)}\,.
\end{eqnarray}

Then by (\ref{NGP}), the Fourier transformation of (\ref{cor21}) and (\ref{cor22}) are 
\begin{eqnarray}\label{G2K}
G^{(1)}(p_1,p_2; k_1,k_2)&=&-\frac{i}{2\pi}\sum_{P,Q}s_{p_{Q_1}}s_{k_{P_2}}\frac{1}{p_{Q_2}-k_{P_2}-i\epsilon}\delta(p_1+p_2-k_1-k_2)\,,\\
G^{(2)}(p_1,p_2; k_1,k_2)&=&\frac{i}{\pi}\sum_{P,Q}s_{p_{Q_1}}s_{k_{P_2}}\frac{1}{k_1+k_2-2\alpha-\chi}\delta(p_1+p_2-k_1-k_2)\,,
\end{eqnarray}
where in (\ref{G2K}) an infinitesimal imaginary part in the denominator arises due to the Fourier transform of the $\theta$ function. 
Moreover, we note that
\begin{equation}\label{dirac}
\frac{1}{p-k-i\epsilon}=\frac{{\cal{P}}}{p-k}+i\pi\delta(p-k)\,.
\end{equation}
On the other hand, since the connected two-photon S matrix contains only a single $\delta$ function, when we apply (\ref{dirac}) to (\ref{G2K}), only the principal part 
contributes to $S^C_{p_1,p_2; k_1,k_2}$. Therefore, we have
\begin{equation}
S^C_{p_1p_2;k_1k_2}=\sum_{j=1}^2 i{\cal{M}}^{(j)}_{p_1p_2;k_1k_2}\delta(p_1+p_2-k_1-k_2)\,,
\end{equation}
with
\begin{eqnarray}
i{\cal{M}}^{(1)}_{p_1p_2;k_1k_2}&=&-\frac{i}{2\pi}\sum_{P,Q}s_{p_{Q_1}}s_{k_{P_2}}\frac{{\cal{P}}}{p_{Q_2}-k_{P_2}}\,,\\
i{\cal{M}}^{(2)}_{p_1p_2;k_1k_2}&=&\frac{i}{\pi}\sum_{P,Q}s_{p_{Q_1}}s_{k_{P_2}}\frac{1}{k_1+k_2-2\alpha-\chi}\,.
\end{eqnarray}
For the connected two-photon S matrix, we can sum over all the permutation terms and obtain a compact form :
\begin{equation}\label{2T}
S^C_{p_1p_2;k_1k_2}=-\frac{\chi}{\pi \gamma} s_{p_1}s_{p_2}\left(s_{k_1}+s_{k_2}\right)\frac{1}{k_1+k_2-2\alpha-\chi}\delta(p_1+p_2-k_1-k_2)\,.
\end{equation}
The final result (\ref{2T}) indeed has the exact analytical structure constrained by the cluster decomposition principle \cite{xf}. 
The only singularities are isolated poles corresponding to one and two-photon excitations in the local system. All principle parts cancel.   

Finally, we sketch the computation for the connected three-photon S matrix. 
we start by computing the six-point Green function $G_3(t'_1,t'_2,t'_3;t_1,t_2,t_3)$. 
Similar to the previous calculation on four-point Green function,
 terms that have non-zero contributions to the six-point Green function can be classified into five types:
$\langle {a}{a^{\dag}}{a}{a^{\dag}}{a}{a^{\dag}}\rangle$,
$\langle {a}{a}{a^{\dag}}{a^{\dag}}{a}{a^{\dag}}\rangle$,
$\langle {a}{a^{\dag}}{a}{a}{a^{\dag}}{a^{\dag}}\rangle$,
$\langle {a}{a}{a^{\dag}}{a}{a^{\dag}}{a^{\dag}}\rangle$,
and $\langle {a}{a}{a}{a^{\dag}}{a^{\dag}}{a^{\dag}}\rangle$. 
Take the last type as an example, we compute its contribution to the six-point Green function is
\begin{eqnarray}
G^{(5)}(t'_1,t'_2,t'_3; t_1,t_2,t_3)&\equiv&(-\gamma)^3\sum_{P,Q}\langle 0| {a}(t'_{Q_1}){a}(t'_{Q_2}){a}(t'_{Q_3})
{a^{\dag}}(t_{P_1}){a^{\dag}}(t_{P_2}){a^{\dag}}(t_{P_3})|0\rangle\nonumber\\
&&\times\theta(t'_{Q_1}-t'_{Q_2})\theta(t'_{Q_2}-t'_{Q_3})
\theta(t'_{Q_3}-t_{P_1})\theta(t_{P_1}-t_{P_2})\theta(t_{P_2}-t_{P_3})\nonumber\\
&=&6(-\gamma)^3\sum_{P,Q} e^{-i\alpha(t'_{Q_1}-t_{P_3})}e^{-i(\alpha+\chi)(t'_{Q_2}-t_{P_2})}e^{-i(\alpha+2\chi)(t'_{Q_3}-t_{P_1})}\nonumber\\
&&\times\theta(t'_{Q_1}-t'_{Q_2})\theta(t'_{Q_2}-t'_{Q_3})
\theta(t'_{Q_3}-t_{P_1})\theta(t_{P_1}-t_{P_2})\theta(t_{P_2}-t_{P_3})\,,
\end{eqnarray}
and then its Fourier transformation
\begin{eqnarray}\label{G5}
G^{(5)}(p_1,p_2,p_3; k_1,k_2,k_3)&=&\frac{3i\gamma}{2\pi^2}\sum_{P,Q}s_{p_{Q_1}}s_{k_{P_3}}\frac{1}{k_1+k_2+k_3-3\alpha-3\chi}
\frac{1}{p_{Q_1}+p_{Q_2}-2\alpha-\chi}\frac{1}{k_{P_2}+k_{P_3}-2\alpha-\chi}\nonumber\\
&&\times\delta(p_1+p_2+p_3-k_1-k_2-k_3)\nonumber\\
&\equiv&i{\cal{M}}^{(5)}_{p_1p_2p_3;k_1k_2k_3}\delta(p_1+p_2+p_3-k_1-k_2-k_3)
\end{eqnarray}
where $P,Q$ are permutations over indices $\{1,2,3\}$. Since (\ref{G5}) contains only a single $\delta$ function, all terms in (\ref{G5}) 
contribute to the connected three-photon S matrix.
Similarly, we calculate the other four types' contributions to $S^C_{p_1p_2p_3;k_1k_2k_3}$ by applying (\ref{dirac}) and keeping only the principal parts. 
The final result is summarized as:
\begin{equation}
S^C_{p_1p_2p_3;k_1k_2k_3}=\sum^5_{j=1}i{\cal{M}}^{(j)}_{p_1p_2p_3;k_1k_2k_3}\delta(p_1+p_2+p_3-k_1-k_2-k_3)\,,
\end{equation}
where
\begin{equation}\label{M1}
i{\cal{M}}^{(1)}_{p_1p_2p_3;k_1k_2k_3}=\frac{1}{4\pi^2}\sum_{P,Q}s_{p_{Q_1}}s_{k_{P_2}+k_{P_3}-p_{Q_3}}s_{k_{P_3}}
\frac{{\cal{P}}}{p_{Q_3}-k_{P_3}}\frac{{\cal{P}}}{p_{Q_1}-k_{P_1}}\,,
\end{equation}
\begin{equation}
i{\cal{M}}^{(2)}_{p_1p_2p_3;k_1k_2k_3}=\frac{1}{2\pi^2}\sum_{P,Q}s_{p_{Q_1}}s_{k_{P_2}+k_{P_3}-p_{Q_3}}s_{k_{P_3}}
\frac{{\cal{P}}}{p_{Q_3}-k_{P_3}}\frac{1}{p_{Q_1}+p_{Q_2}-2\alpha-\chi}\,,
\end{equation}
\begin{equation}
i{\cal{M}}^{(3)}_{p_1p_2p_3;k_1k_2k_3}=-\frac{1}{2\pi^2}\sum_{P,Q}s_{p_{Q_1}}s_{k_{P_2}+k_{P_3}-p_{Q_3}}s_{k_{P_3}}
\frac{{\cal{P}}}{p_{Q_1}-k_{P_1}}\frac{1}{k_{P_2}+k_{P_3}-2\alpha-\chi}\,,
\end{equation}
\begin{equation}
i{\cal{M}}^{(4)}_{p_1p_2p_3;k_1k_2k_3}=-\frac{1}{\pi^2}\sum_{P,Q}s_{p_{Q_1}}s_{k_{P_2}+k_{P_3}-p_{Q_3}}s_{k_{P_3}}
\frac{1}{p_{Q_1}+p_{Q_2}-2\alpha-\chi}\frac{1}{k_{P_2}+k_{P_3}-2\alpha-\chi}\,,
\end{equation}
\begin{equation}\label{M5}
i{\cal{M}}^{(5)}_{p_1p_2p_3;k_1k_2k_3}=\frac{3i\gamma}{2\pi^2}\sum_{P,Q}s_{p_{Q_1}}s_{k_{P_3}}\frac{1}{k_1+k_2+k_3-3\alpha-3\chi}
\frac{1}{p_{Q_1}+p_{Q_2}-2\alpha-\chi}\frac{1}{k_{P_2}+k_{P_3}-2\alpha-\chi}\,.
\end{equation}

As  a check, let $\chi\rightarrow \infty$, only (\ref{M1}) contributes to the connected three-photon S matrix, which agrees with the result in the case of single two-level atom
\cite{ss}, as expected.
Also, according to the cluster decomposition principle, the connected three-photon S matrix
should only contain poles corresponding to the single, two- and three-photon excitations \cite{xf}. 
In (\ref{M1})-(\ref{M5}), in addition to various poles corresponds to single, two- and three-photon excitations, there are also various singularities associated with the principal parts. 
One can actually prove that these principal parts cancel each other when all permutations are summed together. 
A systematic treatment of such analytic properties of the connected $N$-photon S matrix is beyond the scope of this paper  and will be carried out in future works.

\section{Summary and Final Remarks}
To summarize, in this paper, using the input-output formalism, we provide a computation of $N$-photon S matrix in waveguide QED systems. The main result here is the connection between the $N$-photon S matrix and the Green function of the local system. We also discuss the connectedness structure of the S matrix and the Green function, and how such structure can be used to simplify the computation. Our results are applicable independent of the details of the local system's Hamiltonian, and therefore point to some universal aspects of the properties of waveguide QED systems. As a computational tool, the results here lead to a powerful scheme for $N$-photon S matrix calculation.

Aside from waveguide QED systems, understanding the scattering property of a local quantum system coupled to a continuum has been a problem of fundamental importance in many other branches of physics. For example, in condensed matter physics, the transport properties of a quantum dot can also be formulated in a similar fashion \cite{ma}. We therefore expect our development here to be useful beyond waveguide QED systems.

\section{Acknowledgement}
This research is supported by an AFOSR-MURI program, Grant No.
FA9550-12-1-0488. 

\section{Appendix}
Our aim here is to prove the cluster decomposition property of the $N$-photon S matrix (\ref{dSC}).

\begin{figure}
\includegraphics
[width=0.8\textwidth] {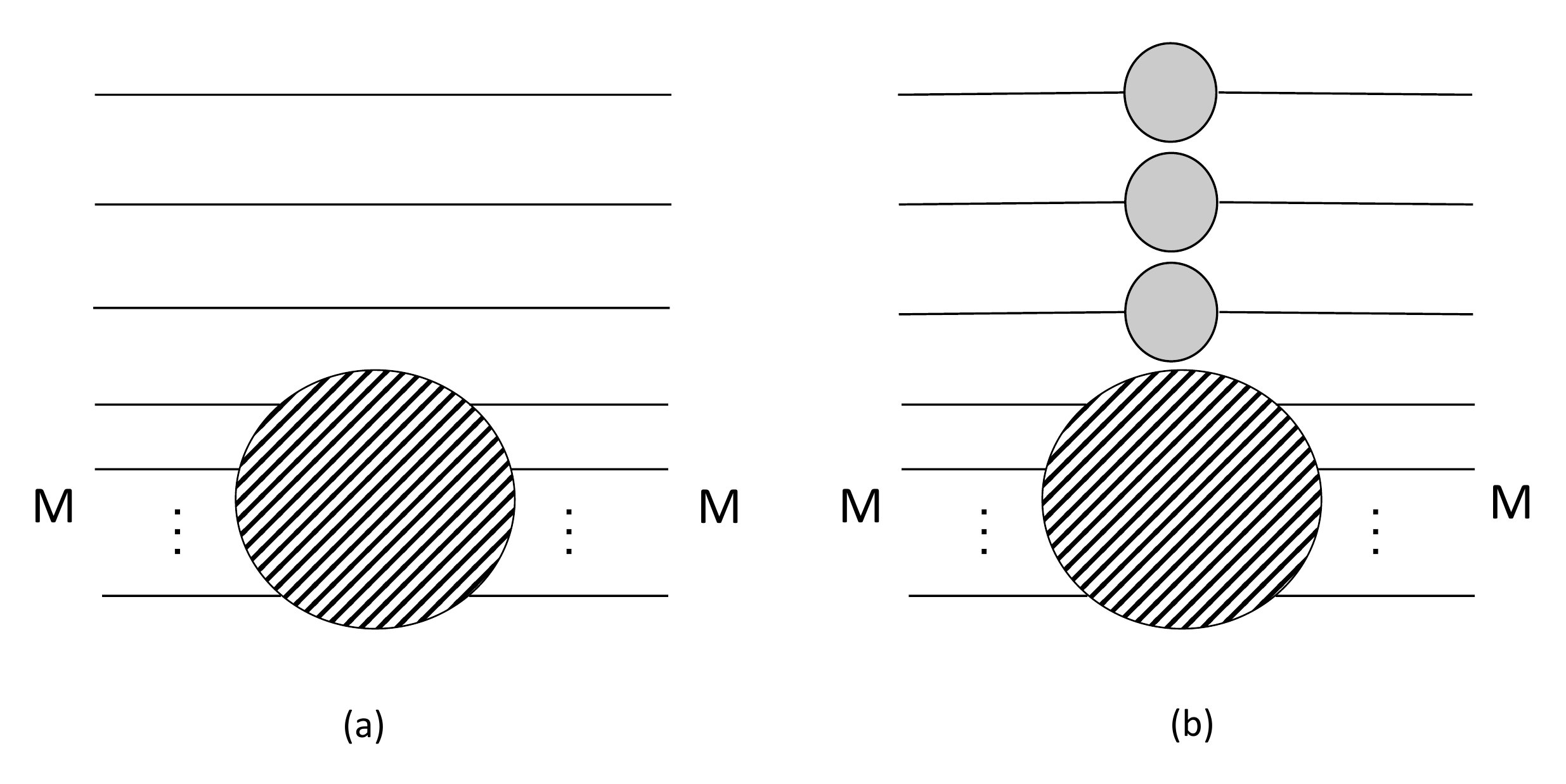} \caption{
(a) A term in (\ref{cenm}) with a $M$-leg sub-piece that contains only the product of the connected $2n$-point Green function with $n\ge 2$.
(b) Same as in (a), but with all the $\delta$ functions replaced by single photon S matrices. This diagram now contains the product of the connected S matrix. 
} \label{fig5}
\end{figure}

We start from (\ref{cenm}), and use the cluster decomposition property of the Green function (\ref{dGC}) to expand every term in the summation in terms of sum over products of connected Green functions and "bare" $\delta$ functions (i.e. the $\delta$ functions that arise explicitly in (\ref{cenm})). Among all these terms resulting from the expansion, we consider the term as represented by the diagram in the Fig.\ref{fig5} (a). This diagram contains a sub-piece with $M$ legs. Within the sub-piece all connected parts are 2$n$-point Green functions with $n \ge 2$. By (\ref{SCGC}) this sub-piece is already in the form of the products of connected S matrices. The rest of this term contains only bare $\delta$ functions. This term obviously arises from the expansion of a term in (\ref{cenm}) containing a $2M$-point Green function.

\begin{figure}
\includegraphics
[width=1.0\textwidth] {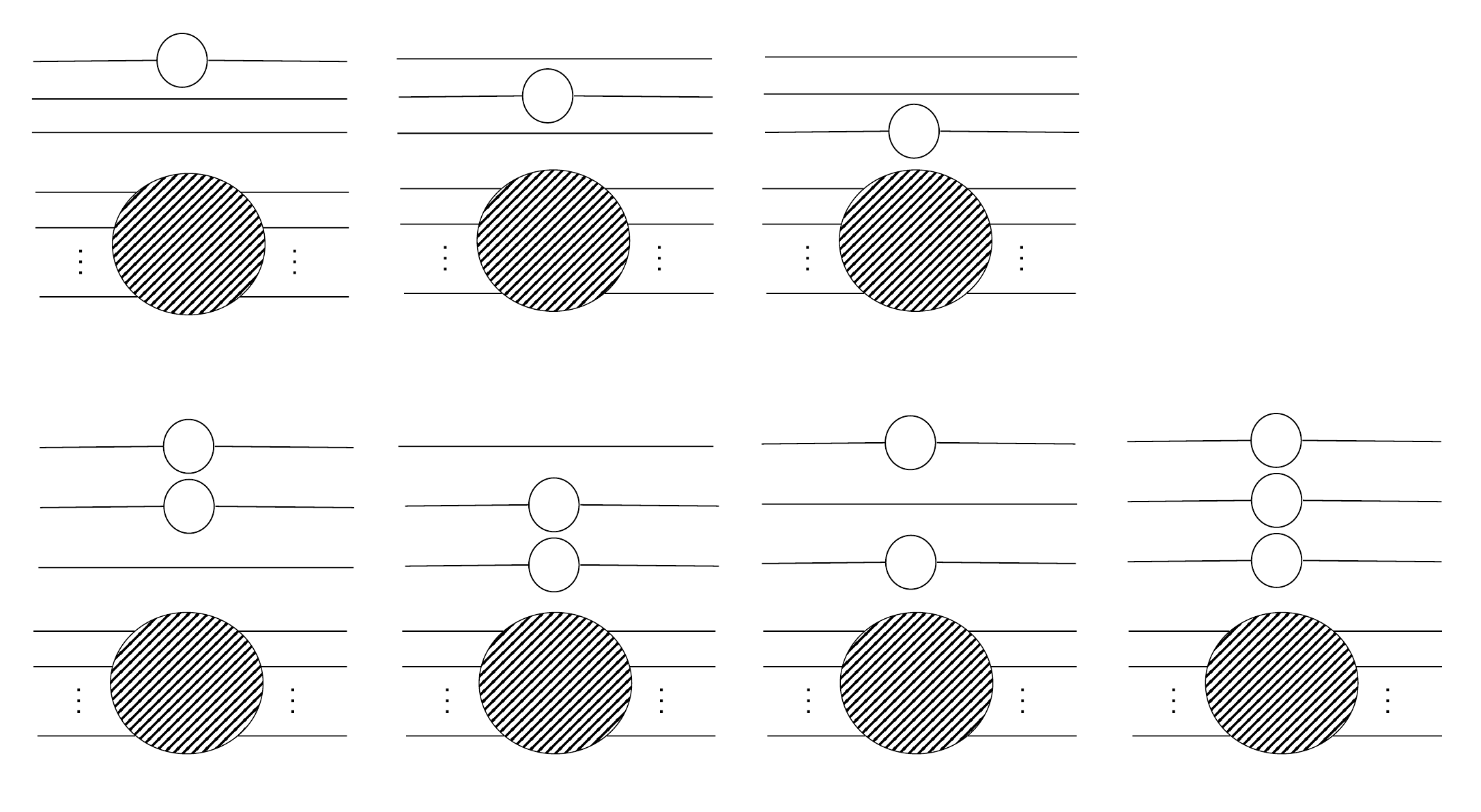} \caption{
All the terms that will be summed over with the term represented in Fig.\ref{fig5} (a) to produce the term in Fig.\ref{fig5} (b).} \label{fig6}
\end{figure}

We combine the term as shown in Fig.\ref{fig5} (a),  with the terms shown in Fig.\ref{fig6}. These terms were chosen from the expansion of the terms in (\ref{cenm}) containing $2M'$-point Green function with $M'>M$. In selecting these terms, we keep the $M$-leg sub-piece identical in all the terms, and choose only terms where the remaining sub-pieces are either bare $\delta$ functions or two-point Green functions. Repeatedly using (\ref{S1G1}), the summation of all these terms then results in a diagram that has the same structure as Fig.\ref{fig5} (a), but with the bare $\delta$ functions all replaced by the single-photon S matrices, as shown in Fig.\ref{fig5}(b). The summation therefore results in a product of the connected S matrices.

Since after the expansion of (\ref{cenm}) in terms of the product of the connected Green's function, every term shows up once and only once in a summation of the form shown in Fig.\ref{fig5} (a) and Fig.\ref{fig6}. We have therefore proved that (\ref{cenm}) can be summed to give the cluster decomposition property of (\ref{dSC}).

\end{document}